\documentclass[%
reprint,
 superscriptaddress,
 amsmath,assume,
 aps,
]{revtex4-2}

\usepackage{graphicx}
\usepackage{dcolumn}
\usepackage{bm}
\usepackage{hyperref}
\hypersetup{ 
  pdfborder={0 0 0},
}
\usepackage{color}
\usepackage{braket}
\usepackage{siunitx}
\usepackage{physics}

\begin{document}

\preprint{APS/123-QED}

\title{Closed-loop measurements in an atom interferometer gyroscope\\with velocity-dependent phase-dispersion compensation}

\author{Tomoya Sato}
\affiliation{Institute of Integrated Research, Institute of Science Tokyo,\\
4259 Nagatsuta-cho, Midori-ku, Yokohama, Kanagawa, 226-8501, Japan}

\author{Naoki Nishimura}
\affiliation{Department of Physics, Institute of Science Tokyo,\\
2-12-1 Ookayama, Meguro-ku, Tokyo, 152-8550, Japan}

\author{Naoki Kaku}
\affiliation{Department of Physics, Institute of Science Tokyo,\\
2-12-1 Ookayama, Meguro-ku, Tokyo, 152-8550, Japan}

\author{Sotatsu Otabe}
\affiliation{Institute of Integrated Research, Institute of Science Tokyo,\\
4259 Nagatsuta-cho, Midori-ku, Yokohama, Kanagawa, 226-8501, Japan}

\author{Takuya Kawasaki}
\affiliation{Institute of Integrated Research, Institute of Science Tokyo,\\
4259 Nagatsuta-cho, Midori-ku, Yokohama, Kanagawa, 226-8501, Japan}

\author{Toshiyuki Hosoya}
\affiliation{Product Development Center, Japan Aviation Electronics Industry, Ltd.,\\
3-1-1, Musashino, Akishima-shi, Tokyo, 196-8555, Japan}

\author{Mikio Kozuma}
\email{corresponding author: kozuma@qnav.iir.isct.ac.jp}
\affiliation{Institute of Integrated Research, Institute of Science Tokyo,\\
4259 Nagatsuta-cho, Midori-ku, Yokohama, Kanagawa, 226-8501, Japan}
\affiliation{Department of Physics, Institute of Science Tokyo,\\
2-12-1 Ookayama, Meguro-ku, Tokyo, 152-8550, Japan}

\date{\today}

\begin{abstract}
Atom interferometer-based gyroscopes are expected to have a wide range of applications
due to their high sensitivity.
However, their dynamic range is limited by dephasing caused by velocity-dependent Sagnac phase shift
in combination with the longitudinal velocity distribution of the atoms,
restricting measurements of large angular velocities.
In this study, we present a method for restoring the contrast deterioration in angular velocity measurements
with interferometer gyroscopes using atomic beams.
Our findings show that by introducing the pseudo-rotation effect with appropriate two-photon detunings
for Raman lights in the interferometer,
it is possible to effectively cancel the Sagnac phase shift 
for all atoms in the velocity distribution of the beam.
Consequently, the contrast is unaffected by the rotation.
Furthermore, we applied this method to an interferometer gyroscope
with counter-propagating atomic beams sharing the same Raman lights.
It is found that the angular velocity of the system can be estimated through the detuning point
where the phase difference between the two interferometers is zero.
This approach ensures that the scale factor of the atom interferometer gyroscope
is independent of the change in the longitudinal velocity distribution
of the atomic beam.
We demonstrate our technique using the interferometer gyroscope of thermal atomic beams of rubidium-87,
achieving a measurement of angular velocity of $\mathrm{{1.0}^{\circ}/s}$
even with an acceleration of 0.68~$\mathrm{m/s^2}$
on a three-axis rotation table.
This simple and robust dispersion compensation method with Raman light detuning
benefits dynamic angular velocity measurements in field applications such as the inertial navigation of vehicles.
\end{abstract}

\maketitle


\section{Introduction}

In the three decades following its inception~\cite{kasevich_atomic_1991,kasevich_measurement_1992},
light-pulse atom interferometry has been actively researched
as a sensing tool across many fields for measuring
acceleration~\cite{lautier_hybridizing_2014,cheiney_navigation-compatible_2018},
angular velocity~\cite{gustavson_rotation_2000,durfee_long-term_2006,savoie_interleaved_2018,avinadav_rotation_2020,gebbe_twin-lattice_2021},
gravity~\cite{peters_high-precision_2001,hu_demonstration_2013,li_continuous_2023},
gravity gradient~\cite{snadden_measurement_1998,mcguirk_sensitive_2002},
fundamental constants~\cite{rosi_precision_2014,parker_measurement_2018},
gravitational waves~\cite{dimopoulos_atomic_2008,canuel_exploring_2018,zhan_zaiga_2020,abe_matter-wave_2021},
dark matter and dark force~\cite{abe_matter-wave_2021,panda_measuring_2024}.
In recent years, the performance of atom-interferometry-based inertial sensors
has reached the level of field applications~\cite{geiger_detecting_2011,barrett_dual_2016,bidel_absolute_2018,bidel_absolute_2020,antoni-micollier_absolute_2024}
owing to their improved sensitivity and accuracy.
Among various applications, these sensors are notably expected to be employed in inertial navigation~\cite{titterton_strapdown_2004}.
Inertial navigation is a method of estimating the position of an individual
without relying on external references such as the global positioning system,
which requires highly accurate angular velocity and acceleration sensors.
The accuracy of conventional inertial sensors, such as
fiber optic gyroscopes (FOGs), has significantly improved
in recent years~\cite{lefevre_fiber_2020,lefevre_fiber-optic_2022,song_advanced_2023};
however, the expected high sensitivity that atom interferometry would bring
is essential for highly accurate inertial navigation~\cite{hogan_light-pulse_2009,fang_advances_2012,narducci_advances_2022}.

For inertial navigation,
the sensors should possess a high dynamic range of measurement,
which is a primary research obstacle in employing atom interferometry in inertial sensors.
In atom interferometry,
angular velocity and acceleration cause shifts in the interference phase
(so-called ``Sagnac effect'' for angular velocity~\cite{sagnac_ether_1913,sagnac_preuve_1913,storey_feynman_1994}).
The magnitude of these phase shifts depends on the velocity of the atoms.
Owing to the longitudinal velocity distribution of atoms,
the phase shift due to the angular velocity or acceleration will also have a dispersion,
resulting in signal loss when the average phase shift is observed.
This effect becomes more intense
as the width of the velocity distribution increases relative to the average speed of the atoms.
One of the most successful methods for extending the dynamic range is
to compress the longitudinal velocity distribution of an atomic beam using laser cooling techniques.
Kwolek {\it et al.}~\cite{kwolek_three-dimensional_2020} produced an atomic beam with a narrow velocity distribution
by extracting laser-cooled and trapped atoms as a continuous beam.
They irradiated these atoms
with lights detuned to the Doppler shift
corresponding to the desired longitudinal velocity.
de Castanet {\it et al.}~\cite{darmagnac_de_castanet_atom_2024-1} developed
an alternative method for an inertial sensor using a cold-atom interferometer.
They extended the dynamic measurement range
by mechanically changing the direction of the laser light within the interferometer
to counteract the effects of the applied acceleration and angular velocity measured using classical sensors.
Although these methods are helpful,
building them compactly is demanding
because they require additional optics and light sources for laser cooling, as well as additional mechanical structures.
Moreover, these additional components may introduce instability.
In addition, achieving high accuracy in measuring and controlling the velocity of atoms is challenging.
Therefore, the instability of longitudinal velocity affects the stability of angular velocity measurements,
particularly because the scale factor is dependent on it.

In this study, we demonstrate a closed-loop phase-dispersion compensation method
to enhance the dynamic range of inertial sensors
using a spatial-domain interferometer.
Importantly, this enhancement is achieved without adding any new elements.
This is accomplished by simply adjusting the two-photon detuning of the Raman lights that compose the interferometer.
Our method involves phase compensation
that is dependent on the time-of-flight of the atoms between lights,
i.e., the longitudinal velocity of the atoms.
The concept of velocity-dependent phase dispersion compensation was proposed and demonstrated
through the precise measurement
of the electric polarizability of a sodium atom~\cite{ekstrom_measurement_1995}.
The phase shift owing to the interaction between the atom and the electric field was compensated using a phase shifter,
which creates electric field gradients,
introducing a phase shift that is inversely proportional to the velocity of the atoms.
Another type of phase shifter utilizing optical prism pairs have been proposed
for use in electrical polarizability measurements~\cite{jacquey_dispersion_2008}.
Gustavon {\it et al.}~\cite{gustavson_rotation_2000} successfully measured
the Earth's rotation rate with the phase shifter, which uses Raman lights that construct atom interferometers;
By temporally sweeping the two-photon detuning of Raman lights
and performing theoretical fitting on the resulting interferometer output,
they determined the value that effectively canceled the Doppler shift
induced by the Earth's rotation.

Here,
we experimentally demonstrated the real-time measurement of angular velocity
using the closed-loop technique.
We
present a detailed calculation of the velocity-dependent phase dispersion compensation
for continuously measuring the angular velocity of a system using dual-beam atom interferometers.
In addition, we demonstrated our method
using an atom interferometer gyroscope (AIG) with dual thermal atomic beams
mounted on a three-axis rotation table.
The experimentally measured scale factor remained
unchanged even when acceleration was applied,
causing the velocity of the atomic beam
to vary during its time-of-flight.
Furthermore, the dynamic range of angular velocity measurements
was successfully extended
through the use of the closed-loop technique. 
Note that a similar technique, using single-photon detuning instead of two-photon detuning of Raman light in our method~\cite{roura_circumventing_2017}, has been proposed to eliminate the effect of the gravitational gradient
in universality of free fall experiments using atom interferometers~\cite{asenbaum_atom-interferometric_2020,ahlers_ste-quest_2022}.

\section{Rotation compensation with the two-photon detuning of Raman lights} \label{sec:theory}

\begin{figure}[h]
\includegraphics[scale=0.45]{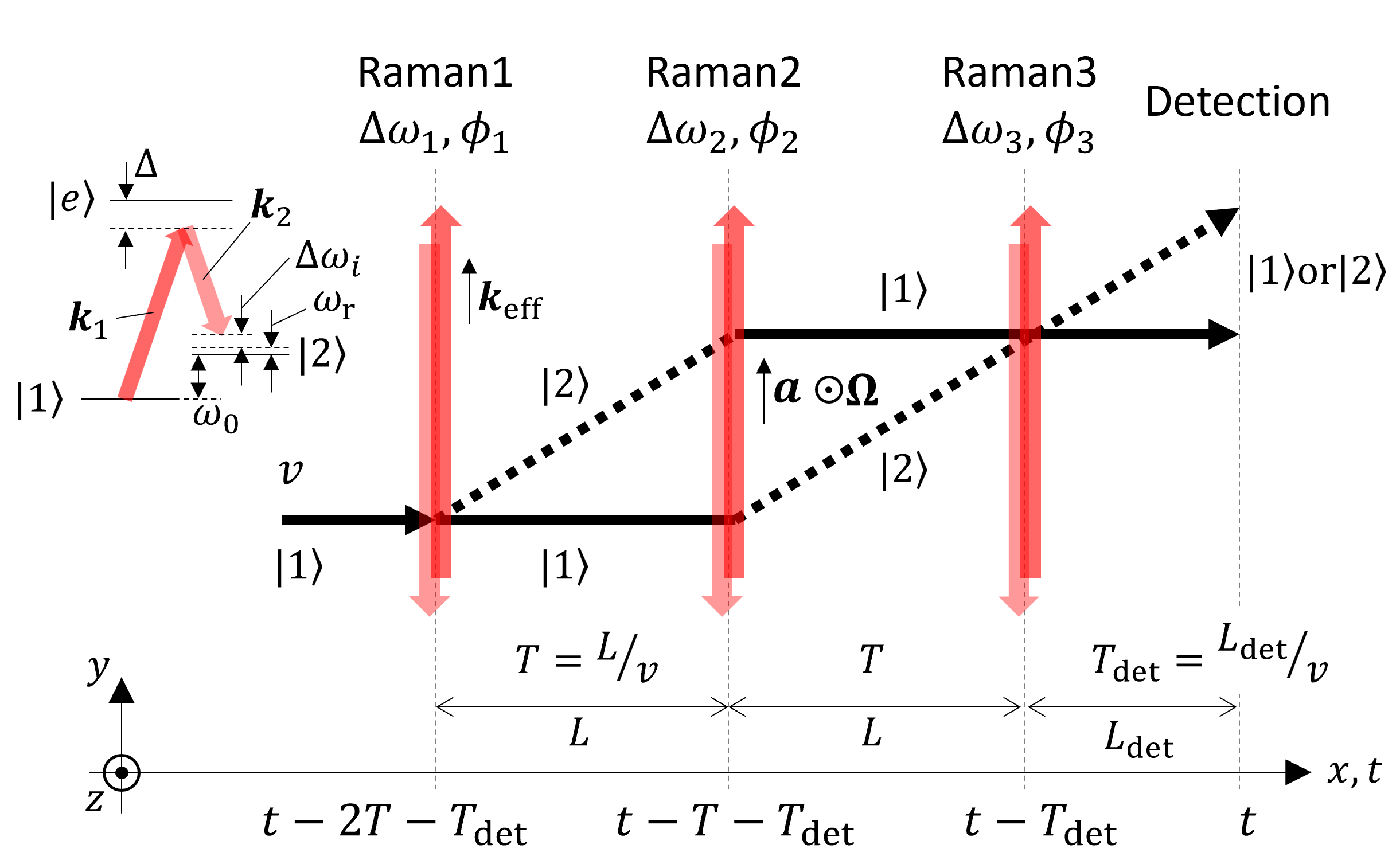}
\caption{\label{fig:figDiagram}
Space--time diagram of a $\pi/2$ -- $\pi$ -- $\pi/2$ Mach--Zehnder-type atom interferometer using Raman transitions.
        }
\end{figure}

Let us consider a $\pi/2$ -- $\pi$ -- $\pi/2$ Mach--Zehnder-type atom interferometer using Raman transitions.
Here, we provide an overview of the Raman transitions and interferometers
(details can be found in references~\cite{berman_atom_1997,gustavson_rotation_2000}).
Figure~\ref{fig:figDiagram} shows the space--time diagram of the single-beam interferometer.
An atom in the ground state $\ket{1}$ and state $\ket{2}$,
whose energy is $\hbar \omega_0$ higher than $\ket{1}$
($\hbar$ is reduced Planck constant), travels at a longitudinal velocity $v$.
When the atom is irradiated with a pair of counter-propagating laser beams,
a two-photon Raman transition between $\ket{1}$ and $\ket{2}$ occurs.
Atoms transitioning to the $\ket{2}$ state through the Raman transition receive recoil momentum
$\hbar(\bm{k_1}-\bm{k_2}) $ from lights,
where $\bm{k_1}$ and $\bm{k_2}$ are the wave vectors of the pair of laser lights.
This causes the atoms in the two states to travel different paths.
The wavelengths of the beams are sufficiently detuned
by $\Delta$ from the excited state $\ket{e}$ to avoid an actual excitation,
and their frequency difference is tuned to $\omega_0 + \omega_{\rm r}$,
where $\omega_{\rm r} = \hbar (k_1 + k_1)^2/(2m)$ is a recoil frequency
and $m$ is the mass of the atom.
An interferometer can be constructed by setting the two-photon Rabi frequency determined
by the intensity of the Raman light
such that the first, second, and final Raman light induce 50$\%$, 100$\%$, and another 50$\%$ transition, respectively.
The population of atoms in the $\ket{2}$ state after passing through the interferometer can be expressed as follows:
\begin{align}
    P_2(\bm{\Omega},\bm{a}) \nonumber \\
    = \frac{1}{2} &\qty{
        1- \cos\qty[ \bm{k_\mathrm{eff}} \cdot (2 \bm{\Omega} \times \bm{v} + \bm{a}) \left( \frac{L}{v} \right)^2 + \phi_{\mathrm{laser}} ]},
    \label{eq:eqPop2}
\end{align}
where $\bm{k_\mathrm{eff}}=\bm{k_1}-\bm{k_2}$ is an effective $k$-vector for the Raman transition,
$\bm{\Omega}$ and $\bm{a}$ are the angular velocity and acceleration vectors of the system,
$\bm{v}$ is the velocity vector of the atom,
$L$ is a spatial separation between Raman lights,
and $\phi_{\mathrm{laser}}$ is an arbitrary laser phase.
This laser phase can be expressed using the phases of each Raman beam $\phi_i (i=1,2,3)$ as
$\phi_{\mathrm{laser}} = (\phi_1 - 2\phi_2 + \phi_3)$.
Here, we assume that the phase outputs of the interferometers are measured simultaneously
using counter-propagating atomic beams that share Raman lights (AIG configuration).
The velocity vector of the left-oriented atomic beam, $\bm{v_\mathrm{L}}$,
exhibits a reversed sign to that of the right-oriented beam $\bm{v_\mathrm{R}}=-\bm{v_\mathrm{L}}=\bm{v}$.
Owing to the orientation dependency of the velocity vector
in the phase term of Eq.~(\ref{eq:eqPop2}),
the rotation term is sign-reversed for both atomic beams,
whereas the acceleration term and arbitrary laser phase remain unchanged.
Thus, by considering the phase difference between two interferometers,
we can obtain the phase shift only due to the rotation,
which can be expressed as: $4\bm{k_\mathrm{eff}} \cdot \bm{\Omega} \times \bm{v} (L/v)^2$.

With the longitudinal velocity distribution of the atoms $f(v)$, Eq.~(\ref{eq:eqPop2}) can be modified as follows:
\begin{align}
    P_2(\bm{\Omega},\bm{a}) \nonumber \\
    &\hspace*{-10mm} = \frac{1}{2} \int_{0}^{\infty} f(v) \times \nonumber \\
    &\hspace*{-5mm}	\qty{ 1- \cos{\qty[ \bm{k_\mathrm{eff}} \cdot (2\bm{\Omega} \times \bm{v} + \bm{a}) \left( \frac{L}{v} \right)^2 +\phi_{\mathrm{laser}} ]}} dv.
    \label{eq:eqPop2Int}
\end{align}
Owing to the finite velocity distribution of the atoms used for the measurement,
the phase of the rotating interferometer will have dispersion.
At high angular velocity,
the interference contrast deteriorates,
limiting the dynamic range of the AIG.
For example,
with the atomic beam of rubidium~(Rb)--87
from the thermal atomic beam source at \SI{100}{\degreeCelsius},
the contrast decreases to $1/e$ at the angular velocity of \SI{0,6}{\degree /s}
with the Raman light separation set at $L=\SI{70}{mm}$.

To restore the decrease in contrast with rotation,
we introduce a velocity-dependent compensation for the Sagnac phase shift
using two-photon detuning of the Raman lights.
The time-dependent phase of Raman lasers $\phi_{t,i=1,2,3}(t)$ can be written as follows:
\begin{align}
    \phi_{t,i}(t) = (\omega_0 + \Delta \omega_i)t + \phi_i,
    \label{eq:eqtDepRamanPhase}
\end{align}
where $\Delta \omega_i$ is the two-photon detuning
from the resonance of Raman transition
in units of angular frequency (see Fig.~\ref{fig:figDiagram}).
As shown in the lower part of Fig.~\ref{fig:figDiagram},
the time at which an atom interacts with each Raman light
depends on its longitudinal velocity.
This implies that atoms receive different phases from the Raman light depending on their velocity
because the phase of Raman light is swept linearly according to the $\Delta \omega_i t$ term in Eq.~(\ref{eq:eqtDepRamanPhase}).
The phase output of the right-oriented atom interferometer is expressed as follows:
\begin{subequations}
\begin{align}
    \Phi_{R} (t,\Omega,a) 
    =& 2 k_\mathrm{eff} \Omega \frac{L^2}{v}
    + k_\mathrm{eff} a \frac{L^2}{v^2} \nonumber \\
    &\quad + (\omega_0 + \Delta \omega_1)(t-2T-T_{\mathrm{det}}) + \phi_1 \nonumber \\
    &\quad - 2\qty[(\omega_0 + \Delta \omega_2)(t-T-T_{\mathrm{det}}) + \phi_2] \nonumber \\
    &\quad + (\omega_0 + \Delta \omega_3)(t-T_{\mathrm{det}}) + \phi_3 \nonumber \\
    =& 2 k_\mathrm{eff} \Omega \frac{L^2}{v}
    + k_\mathrm{eff} a \frac{L^2}{v^2} \nonumber \\
    &\quad- [ (2L+L_{\mathrm{det}}) \Delta \omega_1 \nonumber \\ 
    &\quad\quad\quad\quad- 2(L+L_{\mathrm{det}}) \Delta \omega_2 + L_{\mathrm{det}} \Delta \omega_3 ] \frac{1}{v} \nonumber \\ 
    &\quad+ (\Delta \omega_1 - 2 \Delta \omega_2 + \Delta \omega_3)t + \phi_{\mathrm{laser}},
    \label{eq:eqIntfPhaseL}
\end{align}
where $T = L/v$ is the time-of-flight between the Raman lights with separation $L$,
while $T_{\rm det} = L_{\rm det}/v$ is the time-of-flight
between the final Raman light and the probe light with separation $L_{\rm det}$
(This derivation for a single interferometer can be found also in~\cite{luo_reanalyzing_2021}).
For simplicity, we assume that $\bm{v}$ is aligned along the x-axis, $\bm{k_\mathrm{eff}}$ and $\bm{a}$
are aligned along the y-axis,
and $\bm{\Omega}$ is aligned along z-axis,
i.e. $\bm{v} = (v,0,0)$,
$\bm{k_\mathrm{eff}} = (0,k_\mathrm{eff},0)$,
$\bm{a} = (0,a,0)$ and $\bm{\Omega} = (0,0,\Omega)$.
Similarly, the phase output for the left-oriented atom interferometer (not shown in Fig.~\ref{fig:figDiagram})
can be written as follows:
\begin{align}
    \Phi_{L} (t,\Omega,a) = -&2 k_\mathrm{eff} \Omega \frac{L^2}{v} 
    + k_\mathrm{eff} a \frac{L^2}{v^2} \nonumber \\
    &\quad- [ L_{\mathrm{det}} \Delta \omega_1 - 2(L+L_{\mathrm{det}}) \Delta \omega_2 \nonumber \\
    &\quad\quad\quad\quad\quad\quad\quad\quad + (2L+L_{\mathrm{det}}) \Delta \omega_3 ] \frac{1}{v} \nonumber \\ 
    &\quad + (\Delta \omega_1 - 2 \Delta \omega_2 + \Delta \omega_3)t + \phi_{\mathrm{laser}}.
    \label{eq:eqIntfPhaseR}
\end{align}
\end{subequations}
We assume that $L_{\rm det}$ is the same for both atomic beams.
The first term represents the velocity-dependent Sagnac phase, which causes the dephasing of the interference signal.
Since the third term is proportional to $1/v$,
the phase shift resulting from the Sagnac effect can be nullified
by appropriately setting the sign and value
for the two-photon detunings.
When we set the two-photon detunings
as $\Delta \omega_1 = k_\mathrm{eff} \Omega L$,
$\Delta \omega_2 = 0$,
and $\Delta \omega_3 = -k_\mathrm{eff} \Omega L$,
the relative phase between the two interferometers
becomes zero for an atom
at any given longitudinal velocity:
\begin{align}
    \Delta \Phi (\Omega) =& \Phi_{R} (t,\Omega,a) - \Phi_{L} (t,\Omega,a) \nonumber \\
    =& 4 k_\mathrm{eff} \Omega \frac{L^2}{v} - 2(\Delta \omega_1 - \Delta \omega_3)\frac{L}{v} = 0.
    \label{eq:eqIntfPhaseDiff}
\end{align}
This result shows that closed-loop measurements can be conducted
by controlling the value of the detunings to maintain the condition of $\Delta \Phi = 0$.
Such measurements have three following advantages:

1.~The dynamic range of the rotation measurement using an atom interferometer can be extended
compared to open-loop measurements.
By using closed-loop measurements, the system can be effectively treated as stationary.
As a result, the phase dispersion owing to the velocity-dependent phase shift caused by the rotation vanishes,
thus preventing the dephasing.
This is particularly beneficial for an AIG with a thermal atomic beam,
since it is challenging to implement advanced techniques for suppressing the longitudinal velocity distribution,
such as 3D-cooling in a cold atomic beam~\cite{kwolek_three-dimensional_2020}.

2.~The angular velocity measurement of the system is
independent of the longitudinal velocity of the atoms.
Without the closed-loop operation, the angular velocity is directly derived from the Sagnac phase shift using the following equation:
\begin{align}
    \Omega = \frac{\Delta \Phi (\Omega) v}{4 k_\mathrm{eff}L^2}.
    \label{eq:eqSFOpen}
\end{align}
The precise determination and control of the mean and distribution of the velocity of atoms are challenging,
particularly for thermal atomic beams.
The limited stability of the velocity of the atoms causes errors in the estimation of angular velocities.
With the closed-loop operation,
the measured angular velocity no longer depends on the velocity of the atom, as shown below:
\begin{align}
    \Omega = \frac{\Delta \omega_1}{k_\mathrm{eff}L} = -\frac{\Delta \omega_3}{k_\mathrm{eff}L}.
    \label{eq:eqSFClose}
\end{align}

3.~The acceleration does not contribute
to systematic error in rotation measurement
because the phase shift caused by the acceleration
is common for two interferometers
in Eq.~(\ref{eq:eqIntfPhaseDiff}).

Regarding the advantage of point 3,
it is important to note that the interference contrast diminishes as the acceleration increases.
However, the contrast reduction can be partially compensated
by setting the two-photon detuning of the second Raman light as
$\Delta \omega_2 = - k_{\mathrm{eff}}a{L^2}/\qty[2(L+L_{\mathrm{det}})v]$ (see Eq.~(\ref{eq:eqIntfPhaseL})).
Due to the $1/v^2$ dependence of the phase shift caused by the acceleration,
the compensation is partial
and affected by the longitudinal velocity of atoms.
For example, with the atomic beam of
${}^{87}$Rb from the oven at \SI{100}{\degreeCelsius}
and 
the Raman light separation of $L=\SI{70}{mm}$,
the contrast decreases to $1/e$ at the acceleration of \SI{4}{m/s^2}
which can be improved to $1 \times 10^1$~\SI{}{m/s^2} with compensation.
Since a non-zero $\Delta \omega_2$ works
as a shared acceleration for dual interferometers,
it does not introduce systematic error in the rotation measurement.

The theory of closed-loop measurement for a single-beam atom gyroscope
was proposed by Joyet {\it et al.}~\cite{joyet_theoretical_2012}.
Unlike their approach, our method utilizing the dual atomic beams
offers the advantages above: 2 and 3 under the existence of acceleration.

\section{Experimental demonstration of closed-loop rotation compensation}

\begin{figure}[h]
\includegraphics[scale=0.45]{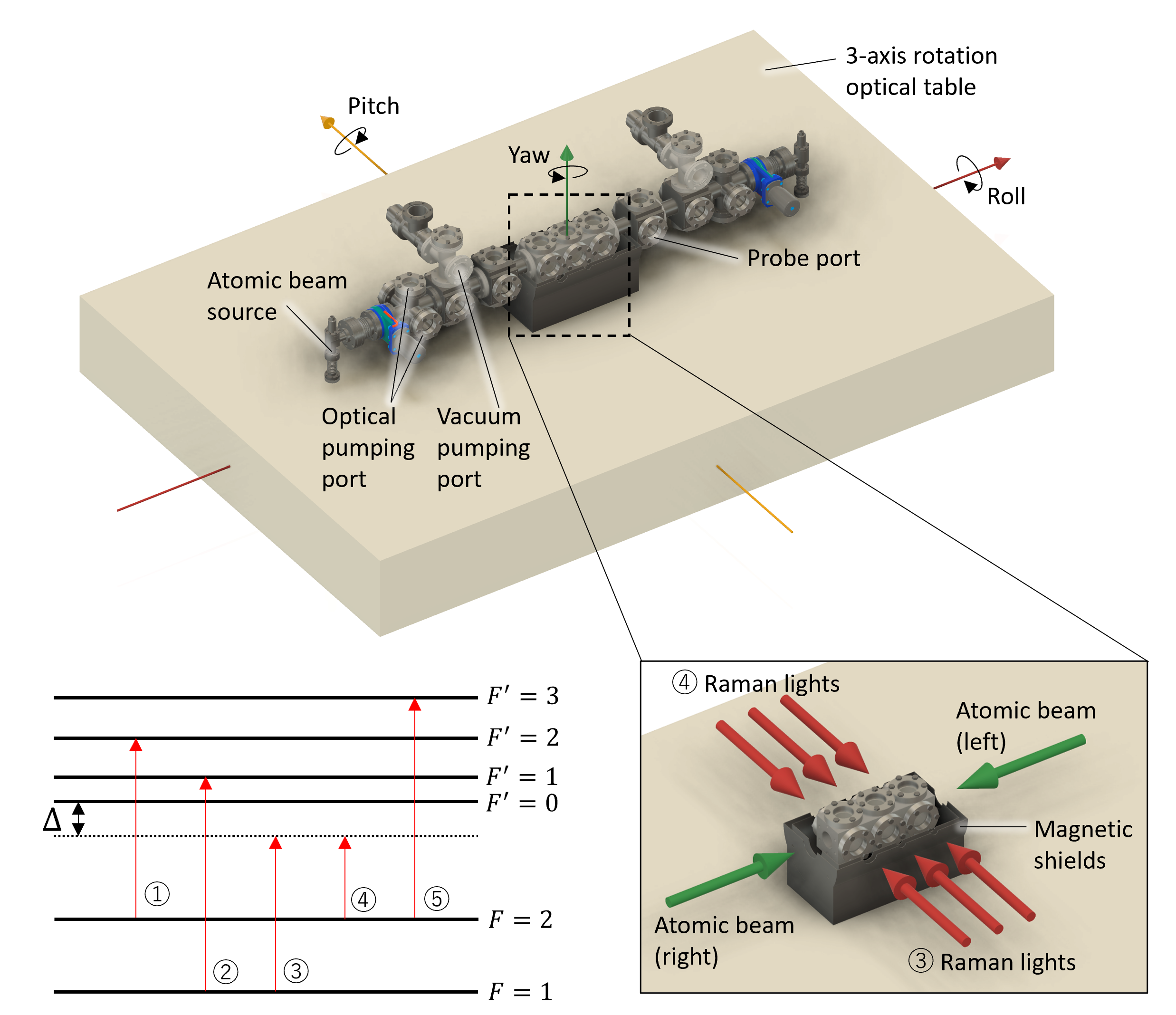}
\caption{\label{fig:figSetup}
        Experimental setup for closed-loop measurements using AIG.
        Legends for one atomic interferometer apply similarly to the other interferometer.
        \raise0.2ex\hbox{{\scriptsize \textcircled{\raise-0.2ex\hbox{\scriptsize 1}}}} Hyperfine pumping light,
        \raise0.2ex\hbox{{\scriptsize \textcircled{\raise-0.2ex\hbox{\scriptsize 2}}}} Zeeman pumping light,
        \raise0.2ex\hbox{{\scriptsize \textcircled{\raise-0.2ex\hbox{\scriptsize 3}}}}
        and \raise0.2ex\hbox{{\scriptsize \textcircled{\raise-0.2ex\hbox{\scriptsize 4}}}} Raman lights,
        \raise0.2ex\hbox{{\scriptsize \textcircled{\raise-0.2ex\hbox{\scriptsize 5}}}} probe light.
        }
\end{figure}

We demonstrated our phase-dispersion compensation method using AIG with thermal atomic beams of ${}^{87} \mathrm{Rb}$.
Figure~\ref{fig:figSetup} shows a schematic of the experimental setup.
We constructed a pair of atom interferometers with counter-propagating atomic beams that shared three Raman lights.
An atomic beam of Rb was extracted from the vapor source at approximately \SI{100}{\degreeCelsius}
through a glass capillary plate with a thickness and hole diameter of \SI{0.5}{mm} and \SI{4}{\micro m}, respectively.
The initial state of atoms was prepared as $m_F\text{=}0$ of $F\text{=}1$ hyperfine ground state
using two pumping laser lights tuned to
the $F\text{=}2$--$F'\text{=}2$ and $F\text{=}1$--$F'\text{=}1$ transitions of $^{87}$Rb.
A $\pi/2$ -- $\pi$ -- $\pi/2$ Mach-Zehnder-type atom interferometer was constructed
using three pairs of Raman lights spatially separated by \SI{70}{mm}.
Each pair of Raman lights was composed of two counter-propagating laser lights,
whose wavelengths are illustrated in Fig.~\ref{fig:figSetup}.
These lights induced the transition between $F\text{=}1$ and $F\text{=}2$ hyperfine ground states
through a Doppler-sensitive Raman transition.
The Raman lights were detuned by $\Delta / 2 \pi = 1.5$\,GHz from the $F'\text{=}0$ state.
The vertical and horizontal waists of each Raman light were \SI{18}{mm} and \SI{190}{\micro m}, respectively.
The powers of Raman lights 1, 2, and 3 were set at 30, 60, and \SI{30}{mW}, respectively,
to achieve the highest interference contrast.
The relative phases of the three pairs of counter-propagating Raman light beams were measured
using the optical beat detection technique, and these light beams were phase-locked to a single radio-frequency reference.
According to the fourth term in Eqs.~(\ref{eq:eqIntfPhaseL}) and (\ref{eq:eqIntfPhaseR}),
we can sweep the phase of the interferometer linearly
by using non-zero two-photon detuning for the Raman light.
This was achieved by inserting an acousto--optical modulator into one of the two lights comprising each Raman light pair.
For rotation compensation,
the two-photon detunings of Raman lights 1 and 3 had equal absolute values with opposite signs.
Moreover, these detunings were dynamically adjusted
such that the difference between the phase outputs of the two interferometers was
equal to the value observed when the rotation table was stationary.
We employed $\Delta \omega_2 / 2\pi = 100$~\si{Hz},
resulting in modulation of the population
in the $F\text{=}2$ state sinusoidally at \SI{200}{Hz}.
Note that, in our experimental parameters,
the $\Delta \omega_2 / 2\pi$ degrading the contrast to $1⁄e$ is estimated to be
$ 6 \times 10^2$~\si{Hz},
which is derived by Eqs.~(\ref{eq:eqPop2Int}) and (\ref{eq:eqIntfPhaseL})
with $L_{\mathrm{det}}$ of 125~mm.
The interference phase can be deduced by
irradiating a probe laser light resonating with the $F\text{=}2$--$F'\text{=}3$ cyclic transition
and performing a lock-in detection of the fluorescence intensity.
In addition to using the Raman transition with $F\text{=}1,m_F\text{=}0$--$F\text{=}2,m_F'\text{=}0$,
which is insensitive to the first-order Zeeman shift,
we also placed a two-layered magnetic shield over the interferometer section to further suppress the second-order Zeeman shift.
A vacuum chamber and magnetic shields enclosing the interferometers, and optical components were constructed
on a three-axis rotation optical table (custom-made by SIGMAKOKI CO., LTD.).
For the yaw axis, rotation in the range of $\pm 15^{\circ}$ at angular velocities up to $1^{\circ}$/s
is available.
For the other axes, tilting is provided in the range of $\pm 4^{\circ}$.

Figure~\ref{fig:figDynRange} shows the angular velocity dependence of the contrast of interference
for right- and left-oriented atomic beams.
The value was normalized to the contrast without rotation of the optical table.
The contrast dependence on the angular velocity with the open-loop measurement
was in good agreement with the theoretical prediction from calculated by Eq.~(\ref{eq:eqPop2Int}).
In the calculation, actual experimental conditions were used:
the mean velocity of the atoms at \SI{100}{\degreeCelsius} escaping from the tube was $v\sim330~\mathrm{m/s}$,
and the arm length was \SI{70}{mm}.
The shifts in the peak positions in the theoretical curve for the open-loop measurement were
due to the phase shift caused
by the two-photon detuning of the second Raman light ($\Delta \omega_2$)
for lock-in detection.
The contrast was maximized at the angular velocity, where the actual rotation cancels out the phase shift.
The slight deviation in the measured peak position from the theoretical curve
in the interferometer with the left-oriented beam
is attributed to the imperfect alignment of the relative angles of the Raman beams against the atomic beam.
For closed-loop dispersion compensation, two-photon detunings for the first and third Raman lights ($\Delta \omega_1$ and $\Delta \omega_3$) were adjusted
such that the closed-loop condition described in Eq.~({\ref{eq:eqSFClose}) was achieved.
We confirmed that the contrast of the atom interferometers was maintained with the closed-loop method,
at an angular velocity of $\mathrm{{0.6}^{\circ}/s}$,
whereas the contrast deteriorated to 1/5 with the open-loop measurement.
We also validated that the contrast did not decay up to an angular velocity of 1$\mathrm{{}^{\circ}/s}$,
which was limited by the performance of the table.

\begin{figure}[h]
\includegraphics[scale=0.60]{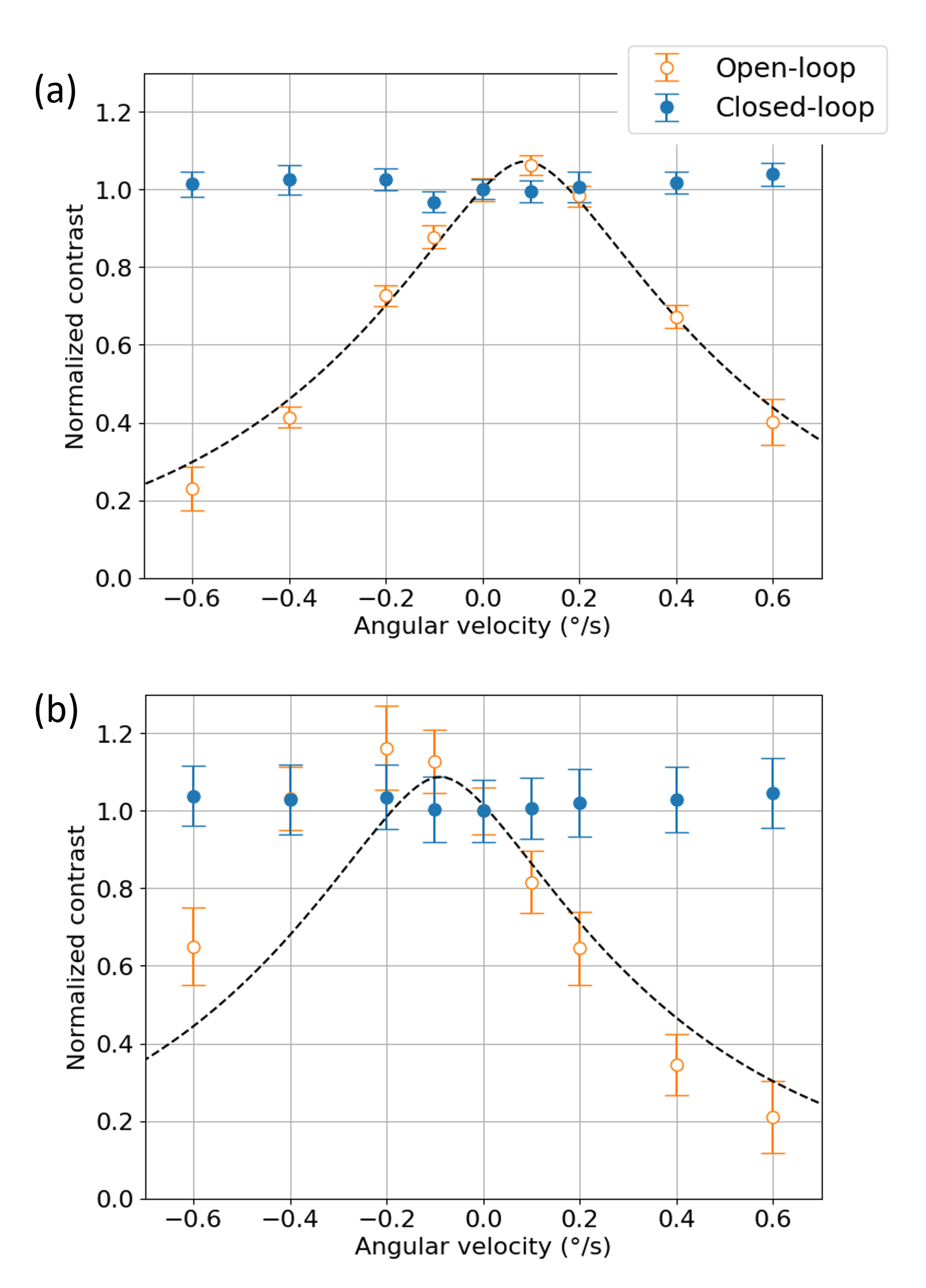}
\caption{\label{fig:figDynRange}
        Angular velocity dependence of the contrast of the interferometer
        with (a) right- and (b) left-oriented atomic beams.
        The values are normalized by the contrast without rotation.
        The dashed lines represent the theoretical estimations
        without the closed-loop rotation compensation for the actual experimental condition.
        Open and filled circles represent the experimental values for open and closed-loop measurements.
        }
\end{figure}

To evaluate the stability of the scale factor, we simultaneously measured the angular velocity of the table
using the AIG and commercial FOG (Exail, blueSeis-3A).
The scale factor stability of the FOG was below 300~ppm,
which had sufficient reliability for validating the concept of our closed-loop technique.
Figure~\ref{fig:figCompFOG} (a) shows the angular velocity measured using the AIG with open- and closed-loop measurements
as a function of the value evaluated using the FOG.
The angular velocity was calculated using Eqs.~(\ref{eq:eqSFOpen}) and (\ref{eq:eqSFClose})
for the open- and closed-loop measurements, respectively.
Figure~\ref{fig:figCompFOG} (b) shows the residual of the linear fitting of the measured data.
The nonlinearity of the scale factor was observed in open-loop measurements.
This could be because the higher-order term was neglected
in the discussion in the previous sections~\cite{bongs_high-order_2006,hogan_light-pulse_2009}.
With increasing angular velocity, the error bars increased,
indicating that interferometer contrast decreased owing to dephasing induced by rotation.
In closed-loop measurement, the linearity of the AIG was maintained at an angular velocity of 1$^{\circ}$/s.
Because the contrast of the interference did not decrease,
the error bar remained small even at a high angular velocity.
The slope of the linear fitting of the data acquired using the closed-loop control was 0.9901.
The scale factor error is 9,900~ppm, significantly larger than
the FOG's 300~ppm.
The cause of this error could be a misadjustment in the Raman light separation or
another systematic error source, which will be addressed in Section~\ref{sec:Discussion}.

\begin{figure}[h]
\includegraphics[scale=0.5]{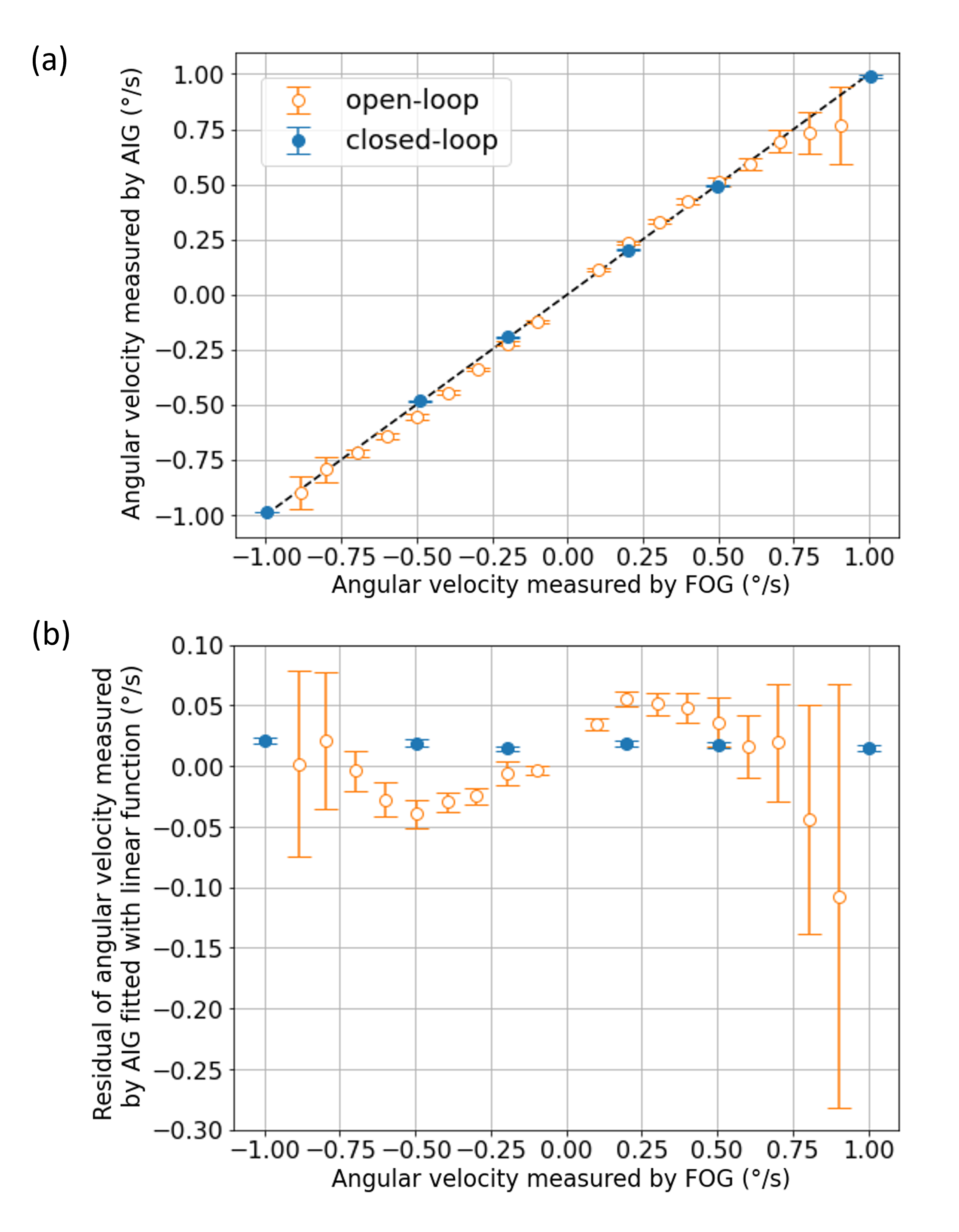}
\caption{\label{fig:figCompFOG}
        (a) Comparison of the measured angular velocity between the AIG and FOG.
        The dashed line represents the fitting result with a linear function for closed-loop measurements.
        (b) Residual errors in the linear fitting of the values obtained from open- and closed-loop measurements.
        In both plots, open and filled circles represent the experimental values for open- and closed-loop measurements.
        }
\end{figure}

The acceleration degrades interference contrast,
whereas it does not cause a systematic error
for measurement of rotational phase shift.
We confirmed this feature by evaluating the angular velocity of the system
while applying constant acceleration using a projective component of Earth's gravity
in the closed-loop measurement, as shown in Fig.~\ref{fig:figSlope}.
In the first measurement, the rotation table was leveled horizontally.
For the second and third measurements, acceleration was applied to the axis to which the interferometer was sensitive.
When the optical table was tilted by 4${}^{\circ}$ along this axis,
it resulted in an acceleration of 0.68~$\mathrm{m/s^2}$.
Consequently, a phase offset of 28.4$\mathrm{{}^{\circ}}$ was observed for both interferometers.
Even if the phases of individual interferometers changed owing to acceleration,
the difference between them was not affected, allowing for accurate angular velocity measurements.
In the fourth and fifth measurements, the interferometer was not sensitive to acceleration,
whereas the velocity of the atom was affected,
causing an error in the scale factor in the open-loop measurement.
For the closed-loop measurement, because the scale factor did not include the velocity of the atoms,
the measured values were unaffected.

\begin{figure}[h]
    \includegraphics[scale=0.5]{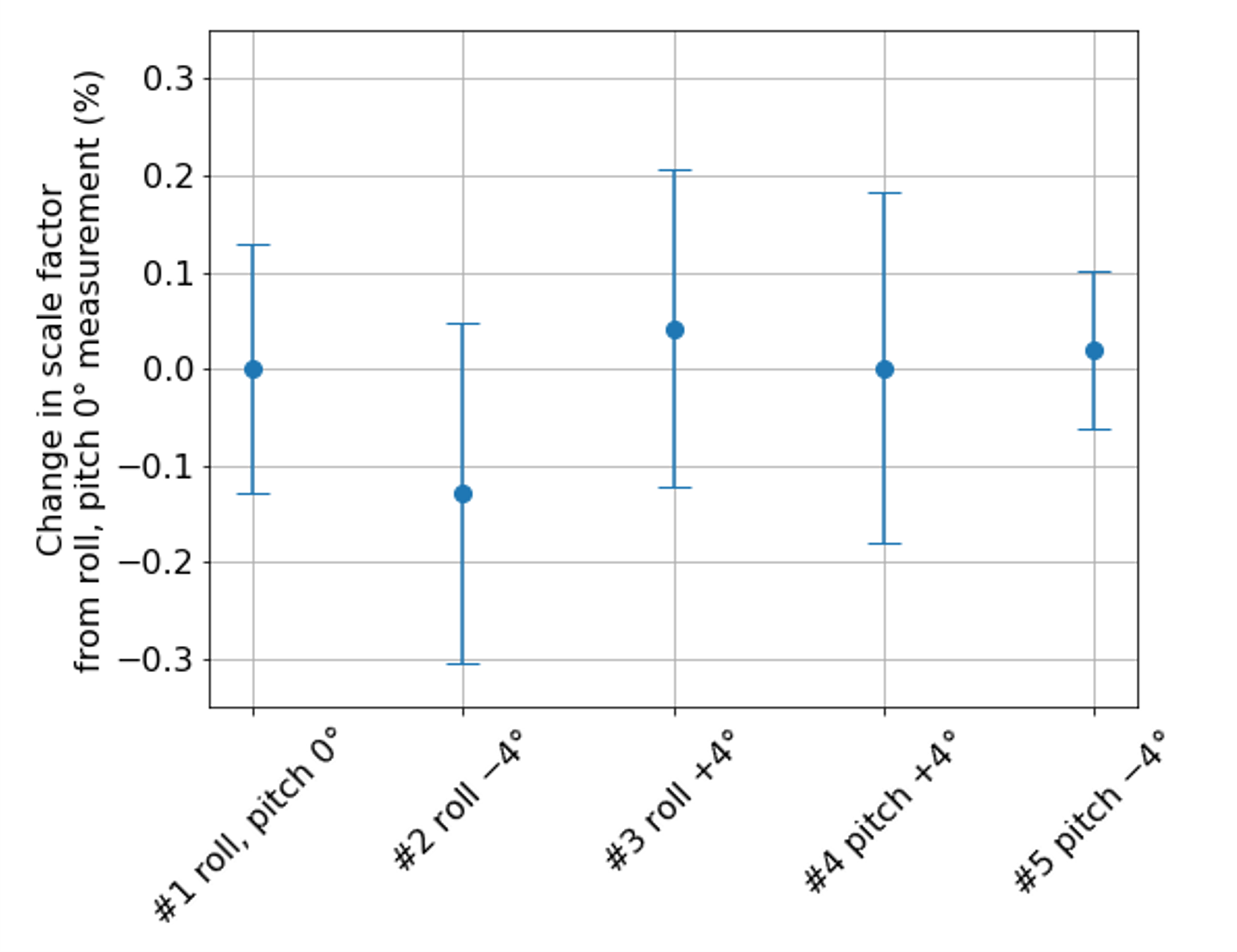}
    \caption{
        Scale factor changes in the rotation measurements with pitch and roll rotation of the table
        in the closed-loop measurements.
        Run $\#$1: Measurements taken with the rotation table as horizontal.
        $\#$2 and $\#$3: Measurements taken with the table tilted by $\pm$\SI{4}{\degree}
        along the axis in the direction of the atomic beam.
        $\#$4 and $\#$5: Measurements taken with the table tilted by $\pm$\SI{4}{\degree}
        along the direction of the Raman beam axis.
        Measured changes in the scale factor are normalized by the value in Run~$\#$1.
        }
    \label{fig:figSlope}
\end{figure}

\section{Discussion} \label{sec:Discussion}

In Section~\ref{sec:theory},
we assumed an ideal setup for the atomic beam gyroscope
to illustrate the core concept of the closed-loop angular velocity measurement.
In this section, we address several systematic errors
arising from imperfections in the experimental setup.


The distance between the last Raman light
and the detection region, $L_\mathrm{det}$, can differ
for right- and left-handed atomic beams.
By substituting $L_\mathrm{det}$ in Eq.~(\ref{eq:eqIntfPhaseL}) and (\ref{eq:eqIntfPhaseR}) with $L_\mathrm{det,R}$ and $L_\mathrm{det,L}$,
Eq.~(\ref{eq:eqIntfPhaseDiff}) is modified as follows:
\begin{align}
    \Delta \Phi (\Omega) =& 4 k_\mathrm{eff} \Omega \frac{L^2}{v} - [ 2L(\Delta \omega_1 - \Delta \omega_3) \nonumber \\
    +& (L_\mathrm{det,R}-L_\mathrm{det,L})(\Delta \omega_1 -2\Delta \omega_2 + \Delta \omega_3) ] \frac{1}{v} = 0.
    \label{eq:eqIntfPhaseDiff_LDet}
\end{align}
Assuming that the two-photon detuning of the first and third Raman light is $\Delta \omega_1 = -\Delta \omega_3 = \Delta \omega$,
the angular velocity can be deduced as:
\begin{align}
    \Omega = \frac{\Delta \omega}{k_\mathrm{eff}L} - \frac{(L_\mathrm{det,R}-L_\mathrm{det,L})\Delta \omega_2}{2k_\mathrm{eff}L^2}
    \label{eq:eqSFClose_LDet}
\end{align}
in the closed-loop measurements.
For the lock-in detection of angular velocity $\Omega$,
$\Delta \omega_2$ must be set to a non-zero value,
which introduces a systematic error into the measurement (the second term in Eq.~(\ref{eq:eqSFClose_LDet})).
However, this systematic error can be compensated with
a single initial calibration of the measured angular velocity,
provided $\Delta \omega_2$ remains unchanged.
Note that $|L_\mathrm{det,R}-L_\mathrm{det,L}|$
can be minimized
by adjusting the position of the probe light
so that the measured angular velocity becomes
independent of changes in $\Delta \omega_2$
under a constant angular velocity.

Next, we discuss the systematic error caused by
the mismatch in the velocity distribution of
counter-propagating atomic beams.
For simplicity, we assume that
the counter-propagating atomic beams
have monochromatic velocities of
$v_{\mathrm{R}}=v,v_{\mathrm{L}}=-\alpha v$,
where $\alpha$ is a positive constant.
Accordingly, Eq.~(\ref{eq:eqIntfPhaseDiff}) should be
modified as follows:
\begin{align}
    \Delta \Phi (\Omega,a) =& 2 k_\mathrm{eff} \Omega \frac{L^2}{v}\frac{\alpha+1}{\alpha}
    + k_\mathrm{eff} a \frac{L^2}{v^2}\frac{\alpha^2-1}{\alpha^2} \nonumber \\
    -& 2 \Delta \omega \frac{L}{v} \frac{\alpha+1}{\alpha} + 2 \Delta \omega_2 \frac{L+L_\mathrm{det}}{v}\frac{\alpha-1}{\alpha} = 0
    \label{eq:eqIntfPhaseDiff_atom-v}
\end{align}
with $\Delta \omega_1 = -\Delta \omega_3 = \Delta \omega$.
From Eq.~(\ref{eq:eqIntfPhaseDiff_atom-v}),
$\Omega$ can be derived as
\begin{align}
    \Omega =
    \frac{\Delta \omega}{k_\mathrm{eff}L}
    + \frac{\alpha - 1}{\alpha} \frac{a}{2v}
    + \frac{\alpha - 1}{\alpha + 1} \frac{L+L_\mathrm{det}}{k_\mathrm{eff}L^2} \Delta \omega_2.
    \label{eq:eqOmega_atom-v}
\end{align}


When $a$ and $\Delta \omega_2$ have finite values,
the estimated value of $\Omega$ will depend on the difference in the velocities
of the counter-propagating atomic beams.
Here, we evaluate the systematic error
in angular velocity measurement
that arises when a slight temperature difference occurs
between the left and right atomic ovens.
In the following simulation,
we used Eq.~(\ref{eq:eqPop2Int})
instead of Eq.~(\ref{eq:eqOmega_atom-v}),
as it accounts for the velocity distribution
of each atomic beam.
We employed an AIG using ${}^{87}\mathrm{Rb}$ atomic beams with an arm length $L$ of 70~mm.
$L_\mathrm{det,R}$ and $L_\mathrm{det,L}$ were set to the same length (125~mm)
because these values could be precisely adjusted using the method stated previously.
The temperatures of the atomic sources were set at \SI{100}{\degreeCelsius} and \SI{101}{\degreeCelsius},
assuming an upper limit for the temperature difference that would be achievable with a commercial temperature controller.
The velocity distributions of the atoms were obtained
using the Boltzmann distribution of the free molecular flow escaping from the tube channel~\cite{ramsey_molecular_1986}.
Under the condition of $\Delta \omega_2 / 2 \pi= 100$~Hz adopted in the current experiment,
a phase shift corresponding to $5 \times 10^{-5}$${}^{\circ}$/s ($9 \times 10^{-7}$~rad/s)
was induced.
Using the recently proposed method of signal acquisition
through phase modulation without frequency sweeping of the Raman light~\cite{kawasaki_analyzing_2024_2},
the systematic error due to finite $\Delta \omega_2$ will nullify
because $\Delta \omega_2$ can be set to zero.
With a finite acceleration of $a=0.68$~$\mathrm{m/s^2}$,
which was applied in this experiment,
a phase shift corresponding to $7 \times 10^{-5}$${}^{\circ}$/s ($1 \times 10^{-6}$~rad/s)
appears.
This systematic error can be eliminated through a real-time correction using Kalman filtering
with acceleration of the system and temperature of sources measured by other sensors.

Finally, we discuss the maximum angular velocity
measurable with our closed-loop method.
As long as the Sagnac phase depends solely on $1/v$,
our closed-loop technique can be applied to
atomic beams with any velocity distribution,
effectively preventing a reduction
in interferometer contrast.
However, deviations from this analysis are expected at
high angular velocities,
where higher-order terms
such as $1/v^2$ become significant.
We numerically simulated the system response under the high angular velocity,
and found that the interference contrast is reduced to $1/e$
at an angular velocity of $3 \times 10^3 {}^{\circ}$/s for our experimental parameters.
This is because the trajectories of the atoms are bent by the Coriolis force,
causing the time-of-flight between the first and second Raman light
to differ from that between the second and third Raman light.

In the discussion of this paper,
we simplified that the direction of atomic beams,
the direction of Raman light propagation and acceleration,
and angular velocity are mutually orthogonal.
When the relationship above is not satisfied,
cross-couplings due to higher-order contributions
will affect the rotational measurement.
Most of the higher-order terms cannot be canceled using our current method.
In addition, cross-coupling due to three-dimensional motion will limit the dynamic range of measurement~\cite{narducci_advances_2022}.
A more detailed evaluation of these effects with theoretical and experimental aspects will be presented in forthcoming papers.

\section{Summary}

The Sagnac phase, which reflects the rotation in atom interferometry, depends on the velocity of the atoms.
Owing to the longitudinal velocity distribution of the atoms,
individual atoms within the interferometer produce varying interference phases,
which results in reduced signal amplitude due to phase dispersion.
We introduce a method to restore the contrast degradation in the AIG.
By setting the frequency detunings to the Raman lights that construct the atom interferometer,
we can induce a velocity-dependent phase shift similar to that of the Sagnac effect.
This introduces a pseudo-rotation effect that can cancel the rotation of all atoms,
even those with a broad distribution of the longitudinal velocity.
Applying this method to an AIG with counter-propagating atomic beams sharing the same Raman lights,
we also observed that the angular velocity of the system can be determined independently of the velocity of the atoms
from the two-photon detunings at which the phase difference between two interferometers becomes zero.
We validated our method using an AIG with thermal atomic beams of $^{87}$Rb 
and closed-loop rotation measurements on a three-axis rotation table.
The contrast of the interference was maintained even at an angular velocity of 0.6$^{\circ}$/s,
which was a significant improvement compared to the 1/5 contrast decrease observed without compensation.
The angular velocities measured using the closed-loop AIG were linearly proportional
to those measured using the commercial FOG up to an angular velocity of 1.0$^{\circ}$/s.
We also demonstrated the robustness of our closed-loop angular velocity measurements against acceleration,
utilizing the projective component of gravity
within the table roll and pitch-tilting in the range of $\pm$4$^{\circ}$.
We performed numerical calculations and found that the possible systematic errors can be eliminated by combination with other techniques and measurements.
With its simple and robust closed-loop mechanism using two-photon detuning of Raman light,
our method holds promise for high-dynamic-range applications such as the inertial navigation of vehicles.

\begin{acknowledgments}
We thank Ryotaro Inoue and Yuichiro Kamino for the helpful discussions.
This work was supported by the Japan Science and Technology Agency (JST), Grant Numbers JPMJMI17A3 and JPMJPF2015.
\end{acknowledgments}

\bibliography{Increasing_dynamic_range_241223_selectedFields}

\end{document}